\documentclass[aps,twocolumn,prb]{revtex4}
\usepackage{graphicx}

\begin {document}
%\wideabs{
\title{Design of Toy Proteins Capable to Rearrange Conformations
in a Mechanical Fashion}

\author{Alexander L.  Borovinskiy*, Alexander Yu.  Grosberg
*\dag}

\affiliation{*Department of Physics, University of Minnesota,
Minneapolis, MN 55455, USA \\ \dag Institute of Biochemical
Physics, Russian Academy of Sciences, Moscow 117977, Russia}
\date{\today}
%\maketitle

\begin{abstract}
We design toy protein mimicking a machine-like function of an
enzyme.   Using an insight gained by the study of conformation
space of compact lattice polymers, we demonstrate the possibility
of a large scale conformational rearrangement which occurs (i)
without opening a compact state, and (ii) along a linear
(one-dimensional) path.   We also demonstrate the possibility to
extend sequence design method such that it yields a "collective
funnel" landscape in which the toy protein (computationally) folds
into the valley with rearrangement path at its bottom. Energies of
the states along the path can be designed to be about equal,
allowing for diffusion along the path.  They can also be designed
to provide for a significant bias in one certain direction.
Together with a toy ligand molecule, our "enzimatic" machine can
perform the entire cycle, including conformational relaxation in
one direction upon ligand binding and conformational relaxation in
the opposite direction upon ligand release.  This model, however
schematic, should be useful as a test ground for phenomenological
theories of machine-like properties of enzymes.

%PACS numbers: 05.40.Fb, 36.20.Ey, 87.15.Cc
\end{abstract}

\maketitle

\section{Introduction}

\subsection{The problem and the context}

Theoretical progress in understanding proteins in the recent years
was concentrated on folding, along with connected questions of
sequence design and evolution (see book \cite{fold_book} and
references therein for the recent overview).  Folding attracts
theorists not only because it is so important for fundamental
biology and for pharmaceutical industry, but also because it is a
robust universal phenomenon.  Vast number of proteins exhibit the
ability to fold, and there is a widely recognized necessity to
understand the physical principle behind the selection of
sequences capable of fast and reliable folding.

In our opinion, there is one more aspect of proteins which is
equally robust and appealing for theoretical analysis in terms of
some minimal model.  We mean here the ability of many proteins to
function in a machine-like fashion through ordered conformational
rearrangement. This is most obvious for motor proteins whose
function is directly related to certain mechanical
(conformational) movements \cite{motor}.  This is also clear for
ion channels whose function is to mechanically move molecules (or
ions) from one place to the other (e.g., across the membrane).
Although less obvious, conformational motions appear to be also
very much at play in proteins whose function is purely electronic,
such as, e.g., electron transfer in bioenergetics or catalysis of
a chemical reaction.  This latter point was first formulated by
McClare \cite{McClare} and independently by Blumenfeld
\cite{Blum_Russian}.   More recently, it was extensively discussed
in the book \cite{Blumenfeld}.  Somewhat different viewpoint on
this subject was also recently presented in the book
\cite{Chernavskii}.

The new experimental data by H. Gruler et al \cite{Gruler} support
the idea of slow conformational relaxation being important for the
operation of enzimatic molecular machines.  More detailed
phenomenological models of enzyme operation based on the concept
of conformational relaxation as a biased diffusion process have
been successfully implemented to interpret the experimental
results \cite{Mikhailov}.  There is now the Data Base of
conformational movements in proteins \cite{database}.

We emphasize two general properties of function-related
conformational movements in proteins.  First, they occur without
significant opening of the dense globular structure.  Viewed in
the context of contemporary folding theories, this property seems
quite exciting.  Indeed, as native globule is pretty dense,
frequently modeled as a {\em maximally} compact self-avoiding
polymer \cite{DENSE}, the inside movements may be expected to be
strongly suppressed.  Counterargument to this suggests that in
fact real protein globule does have certain voids and is not
absolutely dense \cite{VOIDS}.  Nevertheless, the density of a
typical protein globule is similar to that of a polymer melt, for
which reason it may be expected to be extremely viscous if not
altogether glassy.  The observation of significant conformational
movements inside such a dense polymeric conglomerate challenges
theory to offer an explanation.

Second general property we would like to mention is the presence
of some preferred collective degree of freedom - which is almost a
synonym to a functioning device.  For instance, enzymes work in
cycles, and each cycle means a turn around some loop in the
conformational space.  For channel-forming protein, a part of this
loop corresponds to a transported ion moving from one place to the
other, the rest of the loop corresponds to the protein coming
back.  According to the arguments developed a long time ago (see
the book  \cite{Blumenfeld} and references therein), it is
important that there is only one collective degree of freedom
along the loop of function (which, of course, does not rule out
"transverse" fluctuations \cite{fluctuations}).

Importantly, machine-like function is realized well away from
equilibrium conditions.  That means, there is no detailed balance,
and the system moves along the loop in one direction and not in
the other - there is no, and should not be, detailed balance.
This, however, does not rule out the possibility that some parts
of the cycle may present themselves as being the motions along the
same path in opposite directions, like, e.g., a piston moving up
and down in a steam engine.  We shall return to this point later.

The notion of preferred function-related degree of freedom may be
compared in some respects to the concept of reaction coordinate
much discussed in folding studies
\cite{trans_coord,DIRECT_CURRENT}. In both cases, the presence of
transverse fluctuations is important.  In case of folding, this
gives rise to the understanding that, e.g., transition "state" is
not a microstate, determined to atomic details, but rather an
ensemble of (micro)states \cite{Baldwin}.  The preferred
functional degree of freedom must be considered in pretty much the
same way.

It is an exciting question whether these two collective degrees of
freedom, relevant for folding and function, are connected to one
another.  One could even speculate that they may be the same, or
similar to some approximation.  As of today, this question remains
open.  However, we note in passing that turn-around times reported
in modern single molecule experiments on enzymes such as
\cite{Feng_Gai}, on the order of a fraction of a millisecond, are
not drastically different from typical folding times.

In the present paper, our goal is modest, but two-fold.  First, we
want to see, at least for the simplest model, how one can imagine
a collective degree of freedom allowing orderly motion without
opening the compact globule.  Second, we want to design such a
sequence that the globule energy changes in some desirable fashion
while moving along the preferred coordinate. This way, we want to
mimic a molecular spring.

Thus, our work is organized as follows.  We first describe the
model and formulate our problem in a more explicit way for that
model.  We then discuss the possibility to design the native state
conformation, or, better to say, the ensemble of nearly-native
state conformations, in such a way that they can realize the
one-dimensional motion within an almost compact globule.  After
that, we design the sequence capable to fold into such a
functioning conformation.  At the end we study some properties of
thus designed toy protein.

\subsection{A digression: from spherical cow to lattice protein}

We shall work with the toy lattice model of protein.  We
understand that these words will run the emotions high and
negative with many readers, and so we want to answer that from the
very beginning.  Everyone understands that there are no lattices
in biological world, but this argument itself, however obvious, is
too cheap to prove the lattice models useless. Indeed, for
example, everyone laughs at the famous anecdote about a spherical
cow, but at the same time everyone tacitly agrees that the model
of a spherical cow is useful, e.g., to understand the scaling laws
relating animal body mass and the rate of oxygen consumption
\cite{NATURE}.  Thus, the dispute about usefulness, or the lack of
one, for the lattice models in protein studies cannot be resolved
on the level of philosophy, this is the question of specific
purpose of certain studies.  Not entering the details, there are
questions for which lattice models are totally inappropriate (and
may deserve laugh), there are some other questions for which using
lattice models is legitimate.  As we hope to prove by the results,
our present paper belongs to the latter category.

Thus, we use standard toy lattice model in which protein is
represented as a self-avoiding walk of the desirable length.  The
protein changes its conformation by means of elementary moves,
including corner flips, end flips, crankshafts, and null moves
\cite{VS}. The advantage of this move set is that the resulting
system is known to be ergodic \cite{ERGODICITY}.  For this model,
polymer moves by making discrete succession of steps from one
conformation to the next.  Accordingly, the preferred degree of
freedom must be associated with certain a linear (one-dimensional)
succession of conformations, in which every conformation may only
move into either previous or next conformation  of the same group.

\section{Design of conformation}

\subsection{Space of conformations}\label{confspace}

We stated above why the  protein needs to have a selected degree
of freedom for functional work.  Here, we design lattice toy
protein in such a way that its conformations, while remain
compact, can rearrange along one-dimensional linear path in
conformational space.

\begin{figure}[ht]
\centerline{\scalebox{0.7} {\includegraphics{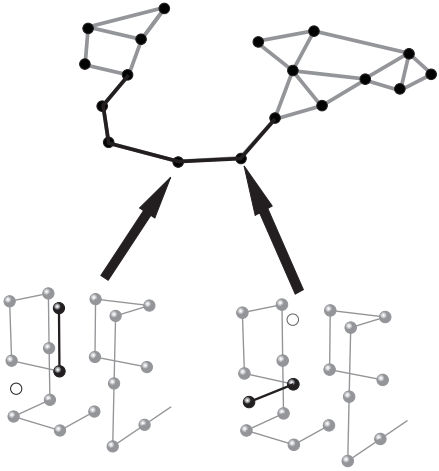}}}
\caption{The conformational space of the lattice polymer can be
visualized as a graph. The nodes represent conformations. Two
nodes are connected if and only if the transition between them is
possible via single elementary move. The linear rearrangement
pathway is shown in bold.}\label{fig:graph}
\end{figure}

The concept of linear path can be easily explained if the
conformational  space of lattice protein is visualized as a graph
\cite{graph1,graph2,graph3}.  In this representation, every
conformation is denoted as a node of the graph, and two nodes are
connected if and only if the transition between corresponding
conformations is possible via single elementary move (see Figure
\ref{fig:graph}).

We are looking for linear paths in the conformational space graph
(CSG).  Obviously, linear path is the succession of such nodes
each of which is connected to exactly two other nodes.  For
swollen, non-compact polymer, a multitude of conformational
motions is possible, the corresponding nodes of the graph have
very many connections, and so none of the swollen conformations
belong to the linear path.  By constrast, in compact conformations
conformational freedom of the polymer is very limited, and there
is a hope to find linear paths.

In order to find one-dimensionally-connected paths of compact
conformations, we examined from this point of view the properties
of CSG of the short lattice polymer.

\subsection{Properties of the space of the compact conformations}

We build on the findings of the work \cite{Foam}.  In that work,
it was shown that placing the polymer inside a restricted size box
makes the conformational space graph disconnected, consisting of
several disjoint pieces, or chambers.  This finding was based on
computer simulation of lattice polymers of various lengths $N$,
each confined in the $3 \times 3 \times 3$ box on the lattice. For
our present purposes, it is important to address another physical
situation, in which $N$ is fixed, while the degree of compactness
of the chain may change.  We achieve this by restricting the
gyration radius of the chain $R_g$ and then looking at various
specific values of $R_g$.  More specifically, this was done in the
following manner.

\begin{figure}[ht]
\centerline{\scalebox{0.85} {\includegraphics{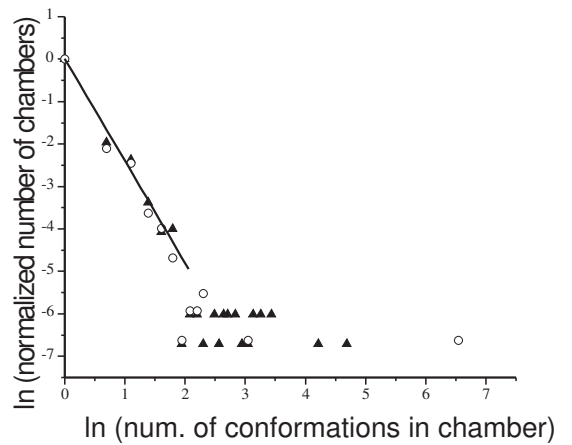}}}
\caption{The dependence of the number of the small chambers in the
CSG on the number of conformations locked in chamber. Two
different CSGs were built limited by $R_g=1.304$ (triangles) and
$R_g=1.305$ (circles). For a comparison, maximally compact
conformations of lattice 18-mer have $R_g=1.2583$. The
conformational space of lattice $18$-mer restricted by $R_g=1.304$
consist of  $1094$ small chambers.  The "infinite" cluster
incorporates $23536$ compact
conformations.}\label{fig:percolation}
\end{figure}

We consider lattice polymer of the length $N=18$.  We start from
maximally compact conformation and allow it to make all elementary
moves consistent with the chain self-exclusion, but possibly (and
necessarily!) violating the compactness.   We accept the
conformation and place it as a node on the graph if and only if
the chain gyration radius in this conformation is less than the
chosen threshold $R_g$. All the accepted conformations are
pictured on the graph as nodes and their connections with all
other accepted conformations are established through exhaustive
search.  Then these new conformations again allowed to make
elementary moves, new conformations are accepted if they do not
exceed the same threshold $R_g$, and the process repeats.  As
regards the limiting value of $R_g$, we choose it experimentally,
and it regulates both the compactness of the conformations and the
number of conformations in the graph.  The graph constructed in
such a manner consists of several disjoint regions \cite{Foam}, or
chambers.  That means, for two conformations, which  belong to
different chambers, there is no sequence of elementary moves,
which transforms one of them into another without breaking the
restriction on $R_g$.  Thus, the procedure must be repeated
starting from different maximally compact conformations to list
all the chambers of the graph.  The number of conformations in
different chambers varies, reflecting the distribution of the
clusters in the bond percolation problem (in this case, we deal
with percolation in conformation space \cite{Foam}).  Figure
\ref{fig:percolation} shows the dependence of the number of the
small chambers in the conformational space graph on the number of
conformations locked in chamber.  In terms of the underlying
physical idea, this figure is similar to the result reported
earlier in the work \cite{Foam} for a different model.  In the
present model, we vary $R_g$ at the fixed number of monomers $N$.
In the work \cite{Foam}, the similar alleged percolation in the
conformational space was controlled by changing $N$ while locking
the polymer inside the $3 \times 3 \times3$ cube on the lattice,
which of course implies fixed $R_g$.

\begin{figure}[ht]
\centerline{\scalebox{0.85} {\includegraphics{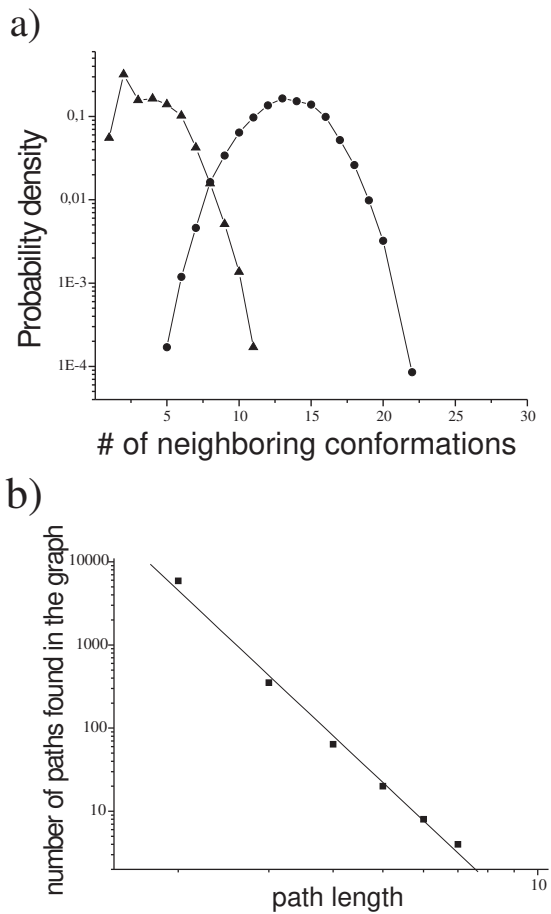}}}
\caption{ (a) The distribution of the connectivities of the nodes
within "infinite" cluster of the conformational space graph of the
lattice 18-mer restricted by $R_g=1.304$ (triangles) compared with
one for the  non-restricted 18-mer (circles). The peak of the
distribution corresponds to the nodes incorporated into the linear
paths.  (b) The distribution of the linear paths in the "infinite"
cluster by length.  The longest path found consist of $9$
elementary moves.}\label{fig:connectivity}
\end{figure}

Further, we established the  connectivities of the nodes which
belong to the largest chamber in the CSG.  The distribution of the
connectivities of the nodes is shown in Figure
\ref{fig:connectivity}a, curve 1. It is compared with the
distribution of the numbers of neighboring nodes for the same set
of conformations, but not restricted with the $R_g$-condition
(curve 2).  Curve 2 is indistinguishable from the binomial
distribution (which does not contradict the idea that
non-restricted CSG is a small-world network \cite{smallworld}).
The CSG built under the restriction on the values of $R_g$ of
toy-protein conformations (curve 1) is significantly different,
one can imagine this graph as a percolation cluster of the bond
percolation problem on the lattice with the topology of the
small-world network \cite{percolation}. For the small values of
$R_g$ (weakly connected cluster) the peak of the distribution
corresponds to the graph nodes connected to only two neighbors.
The sharp peak on the curve 1 corresponding to the poorly
connected conformations can be easily explained.  The change of
the geometry of the voids in the bulk of the protein globule is
possible only via small number of local moves, because the
excluded volume effect is very strong in compact conformations.
Such subtle conformational moves do not affect significantly the
value of the $R_g$ of the protein chain, whereas opening of some
loop on the surface of the globule leads to the increase of the
$R_g$.  On the other hand, the majority of the conformational
moves accessible to the unrestricted protein chain occurs on the
surface.  Accordingly, the limiting of $R_g$ from above forbids
surface moves but does not restrict the changes in the bulk of the
globule.  That is why for small limiting $R_g$ values the sharp
peak of the distribution rises at the poorly connected
conformations.

Thus, the study of $R_g$-restricted conformations of the $18$-mer
provides an example in which there is the peak of connectivity
distribution which corresponds to the graph nodes with
connectivity $2$.  These are desired conformations forming linear
paths.  However, most of these linear paths are rather short and
consist of only few elementary moves.  The distribution of the
lengths of the paths is shown in Figure \ref{fig:connectivity}b,
it is exponential distribution.

Thus, our conclusion so far is this.  Long linear paths exist, but
they are not very common, they are rare.  In this sense, the
situation is similar to that of selection of sequences for toy
proteins capable of folding \cite{fold_book}.  In sequence
selection case, the "good" sequences are exponentially rare among
the random ones.  The goal of sequence design, or selection, is to
fish them out.  Similarly, our goal now is to identify the rare
conformations which are connected by long linear paths in the
conformation space.  In order to do that, we need to understand
better the local geometry of conformations belonging to the linear
paths.  It cannot be done for the chain length of $N=18$ which is
too small.  It is also small compared with typical protein
lengths, about one hundred or more in average. Thus, we repeated
the same procedure of CSG mapping for the toy-protein of the
length $N=125$.  Of course, in this case no exhaustive enumeration
of conformations is possible, and so we performed random sampling
instead.

We started from the  limiting value $R_g=2.44949$, which is equal
to the  gyration radius of the maximally compact conformation. In
this case, no conformational movement is possible, and, therefore,
CSG consists of as many disconnected nodes as the number of
compact conformations.  We now increase $R_g$ by a very small
amount.  By several attempts, we choose the step of $R_g$
increment in such a way that after one step conformation space
remains disintegrated, non-ergodic, consisting of disconnected
chambers \cite{Foam} most of which contain only few conformations.
We then increase $R_g$ step by step, every time mapping the newly
obtained parts of conformation space on the graph, as described
before.  When $R_g$ approaches $2.46$, the percolation transition
occurs and there appears an "infinite" cluster.  We mapped into
the graph $1000000$ conformations from this cluster and searched
for the long linear paths on the graph, as it was done in case of
$18$-mer.

\begin{figure}[ht]
\centerline{\scalebox{0.85} {\includegraphics{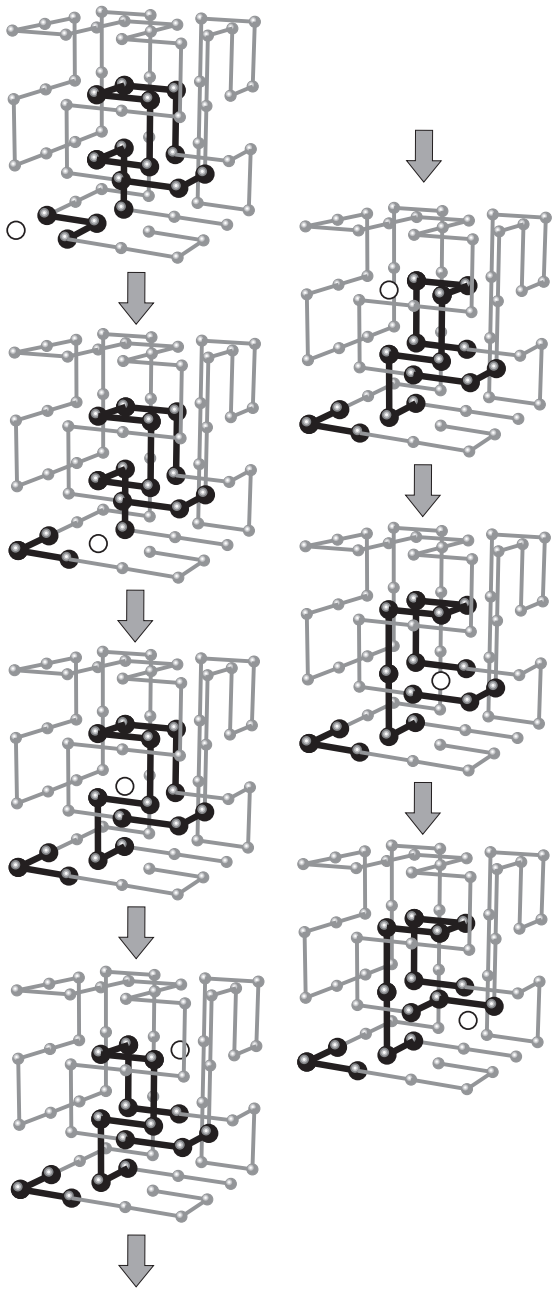}}}
\caption{ A typical realization of the linear rearrangement path
of the compact conformation of the lattice $63$-mer.  The vacancy
travels through the bulk of the conformation in a set of
subsequent corner flips.  The bonds which participate in the
conformation rearrangement are shown in bold color.  The overlap
between conformations on the opposite ends of the pathway
comprises $Q^{1,7}=58$, whereas maximal number of contacts
$Q^{max}=79$.}\label{fig:path}
\end{figure}

The longest paths we found for the $125$-mer are organized as
follows.  The conformations along the path transform into one
another in a set of subsequent corner flips in the bulk of compact
conformation.  When the first corner flip occurs, the vacant site
opens producing the opportunity for another monomer to move. Then
the similar process repeats many times, leading effectively to the
vacancy traveling through the globule, finding the new corner
flips on its way.  These subsequent transformations are
demonstrated in Figure \ref{fig:path} which for the simplicity of presentation
shows a smaller polymer, namely $63$-mer.

\subsection{Design of the pathway in conformation space}
\label{switches}

Now, when the mechanism of rearrangements within the compact
conformation for the toy protein is clarified, we can propose the
effective and straightforward approach to design the linear path
of compact conformations.

The algorithm includes two main stages.  First, we arrange the
switching elements along the path.  Second, the rest of the
polymer is computationally designed such that it is (almost)
maximally compact and contains the necessary set of switching
elements.  As a result, we obtain one-dimensional path of compact
conformations of the lattice toy protein.

\subsubsection{Building the switching elements.} We build
the linear path using flipping corners as switching elements.

\begin{figure}[ht]
\centerline{\scalebox{0.85} {\includegraphics{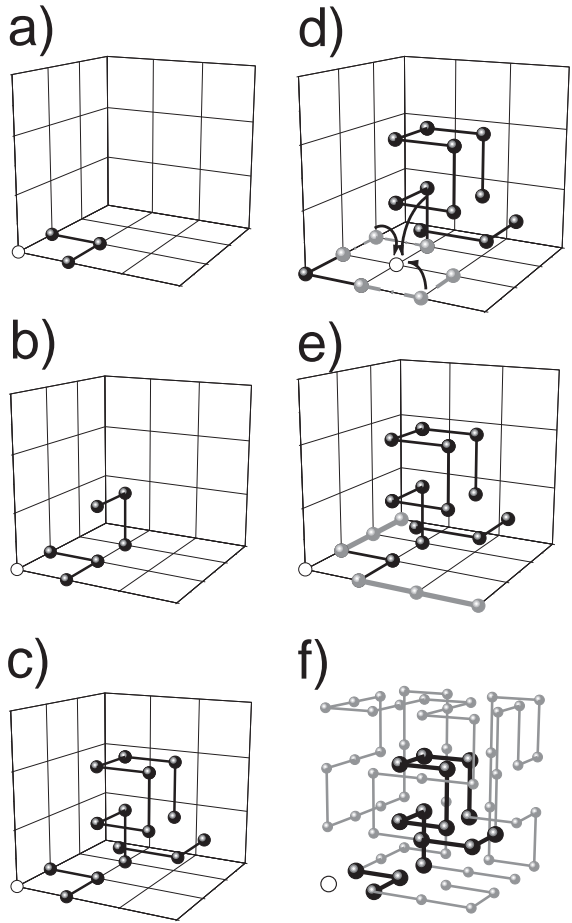}}}
\caption{ The steps of the design of the linear rearrangement
pathway.  (a) The initial position of the vacancy and the first
flipping corner  are dawn on the lattice; (b,c) the other
switching elements of the path are added; (d) possible parasitic
flips, which may later appear on the stage of the design of the
whole conformation, are shown in grey; (e) to prevent parasitic
flips additional elements of the conformation surrounding the
switching elements of the path are added (shown in yellow);(f)
conformation design completed.}\label{fig:design_of_path}
\end{figure}

In the beginning of the procedure we have an empty cubic lattice.
We start from choosing initial position of the vacancy (Figure
\ref{fig:design_of_path}a, node $(0,0,0)$).  The first flipping
corner is drawn on the lattice in such a way that when it flips,
the corner and the vacancy exchange their positions.  The edges
forming next switching element should be drawn in such a way, that
when it will flip, vacancy will hop to the former corner position
opening the way for the next switching element.  In the Figure
\ref{fig:design_of_path}b two subsequent flipping corners are
shown.  After they both will move, the vacancy will take the
position $(1,2,1)$.  Other flipping corners are drawn subsequently
as many times as needed (Figure \ref{fig:design_of_path}c).

At the same time as the  switching elements arranged on the
lattice we can control the linearity of the path.  Every time when
the new flipping corner is set on the lattice there are several
ways to locate it relative to the vacancy position.  We choose
only one location, but the problem is that the other flipping
corners may appear later, on the stage of the design of the whole
conformation (see Figure \ref{fig:design_of_path}d).  To prevent
appearance of such parasitic switches we draw additional pieces of
the conformation surrounding the pathway sites, as it is shown in
Figure \ref{fig:design_of_path}e.

\subsubsection{Building the rest of conformation.}  We now have the set
of switching elements, which are supposed to be just the
disconnected pieces of the polymer.  We have to find now the rest
of the polymer, such that it fills (almost) completely the whole
volume of the cube, and connects all switching elements into a
linear chain.  For this purpose, we developed computational method
which is the modification of the approach proposed by Ramakrishnan
et al \cite{Conf-des} to generate maximally compact conformations.
Here, we describe our modified algorithm.

\begin{figure}[ht]
\centerline{\scalebox{0.8} {\includegraphics{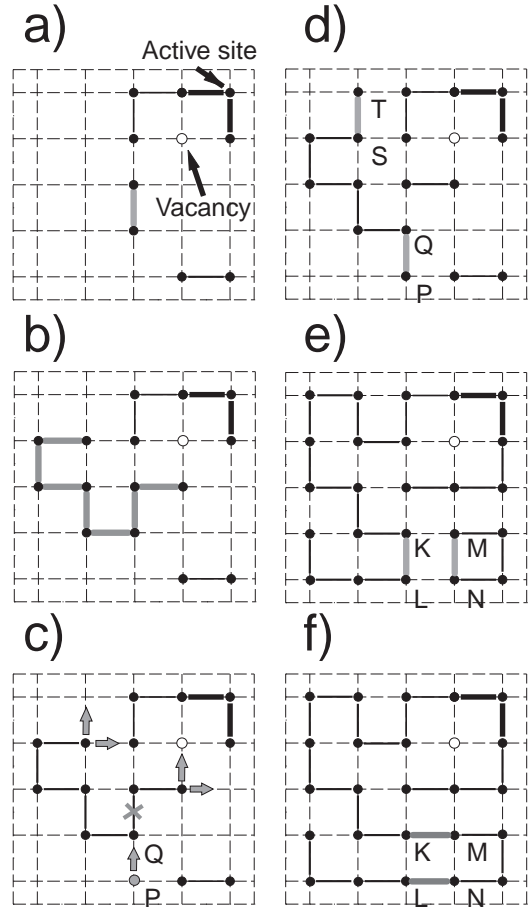}}}
\caption{ The schematic representation of the application of the
algorithm \cite{Conf-des} for the purposes of the design of the
conformation incorporating the rearrangement pathway.  For the
simplicity the steps of the algorithm are shown in two dimensions.
(a) New edges randomly placed  on the lattice; (b) Edges form
subchains; (c) At some step edges can not be placed randomly; (d)
Number of edges increases by one after applying two matching
procedure; (e) Linear subchains an loops; (f) Subchains
united.}\label{fig:filling_volume}
\end{figure}

The conformation design starts with placing on the lattice the new
edges connecting randomly chosen neighboring vertices (Figure
\ref{fig:filling_volume} a,b).  This process soon brings us to the
state, where some vertex cannot be connected to its randomly
chosen neighbor, because the neighbor already has two edges
incident on it (Figure \ref{fig:filling_volume} c). In this case,
special procedure called \emph{two-matching} is applied
\cite{Conf-des}.  During this and the following steps of the
algorithm, some randomly chosen edges can be removed from the
lattice and changed by others. However, we impose the condition
that the edges forming switching elements \emph{cannot be removed}
at any step  of conformation design.

\emph{Two-matching} starts from picking up the vertex $P$, which
is either not connected, or has only one incident edge.  Then its
neighbor $Q$ is chosen randomly as an opposite end of the new
edge. If $Q$ belongs to the linear subchain, then the ends of this
subchain are checked for the possibility to be connected with
their neighbors. If it can be done, a new edge (the edge $ST$ at
the figure \ref{fig:filling_volume}b) is added.  One of the edges
incident on $Q$ is replaced by the edge $PQ$.  Thus, in this
procedure, the number of edges increases by one (Figure
\ref{fig:filling_volume}d).  If, on the other hand, the vertex $Q$
belongs to the looped subchain, one of its incident edges is
removed and replaced by $PQ$.  So, the loop is broken, and the
total number of edges on the lattice remains unchanged. Typically,
as a result of the work of this procedure, we obtain several
looped and one linear subchain packed into the cubic lattice
(Figure \ref{fig:filling_volume}e).

Now, the subchains should be merged into one chain.  This is
achieved in the following way.  Suppose four neighboring vertices
$(K, L, M, N)$ form a square.  The connecting edges $(K,L)$ and
$(M,N)$ belong to the different subchains.  Excluding these edges
and including instead $(K,M)$ and $(L,N)$, we would have merged
subchains (see Figure \ref{fig:filling_volume}f).  Such an
operation is known as \emph{ patching}.  Each patching operation
transforms a pair of subchains into one subchain.  The edges to be
involved in patching are chosen randomly. The process is stopped
when there is no more and no less than one linear chain on the
lattice, which is the desired polymer conformation.

In the original work \cite{Conf-des}, this method was applied to
generate maximally compact lattice conformations.  It is worth
repeating that for our purposes we use this method starting from a
complicated lattice which is the cube minus elements chosen for
switching elements.

Generally, it is possible to use also crankshafts and flipping
ends to design the switching elements.  One just needs to forbid
all the states of these elementary moves but two.  This should be
done by placing on the lattice additional edges (as it was done to
prevent parasitic corner flips), which restrict the extra states
of these moves. However, the end flip can be used only twice as
switching element, because there are two ends of the chain.  As
regards the crankshaft move, it needs two vacant sites to make
switching possible, which means the conformation in question
should be slightly less compact.  For these reasons, we use only
corner flips as switching elements in this work.

\section{Sequence design}

After we have chosen conformations, we should find the sequence
which fits the target conformations with low energies. For this
purpose the sequence of the model protein is annealed and Monte
Carlo optimization in the sequence space is performed.  The
details of the algorithm are previously published \cite{seq-des}
(see also review article \cite{Grosberg1} and references therein
for further details).  It must be emphasized that we plan to work
with the sufficiently large set of monomer species; in fact, we
shall even use the so-called Independent Interaction Model, in
which the number of distinct species is as large as the number of
monomers in the chain.  This allows us to avoid difficulties well
known in the case of sequence design for the two-letter
heteropolymers, such as the $HP$-model \cite{Yue}.  In the context
of the present work, sequence design method had to be modified in
two respects. First of all, we need not only one target
conformation to have a low energy, as it is typically assumed in
protein folding simulations. We need the whole family of
conformations - all conformations belonging to the rearrangement
path - to have distinctly lower energies than all other states.
Second, the more ambitious goal is to design  the sequence in such
a way that moving of the system along its pathway changes energy
in an orderly fashion.  Since our rearrangement path is
one-dimensional, we can pretend energy to increase monotonically
along this path, in this sense making our toy protein a model of a
molecular spring.

Let us consider these two aspects of sequence design one by one.

\subsection{Sequence with multiple ground states}

The model protein is determined by the set of the coordinates of
the monomers ${\cal C}=\{{\vec r}_I\}$ and the sequence of
monomers $seq=\{s_I\}$ (the species $s_I \in \{1,2,...q\}$ denote
the identity of each monomer, $q$ is the total number of species),
index $I$ counts monomers along the chain.  The Hamiltonian is
written as follows
\begin{equation}
{\cal H}({\rm seq}, {\cal C})=\sum_{I<J}^N B_{s_I s_J}\Delta({\vec
r}_I-{\vec r}_J) \ ,
\end{equation}
where the energy  of the conformation is determined by the matrix
of species-species energies $B_{i,j}$ for the contacting monomers
and function $\Delta({\vec r}_I-{\vec r}_J)$ is defined such that
it is equal to $1$ if monomers $I,J$ are lattice neighbors and $0$
otherwise. We use independent interactions model (IIM) \cite{IIM}
and Miyazawa-Jernigan (MJ) \cite{MJ} matrices for Monte Carlo
simulations in this article.

In our approach the goal is  to design  sequences for which the
whole set of conformations $\{{\cal C}_k\}$ have energies
sufficiently below that of the rest of conformation space.  Of
course, our candidates for the target states $\{{\cal C}_k\}$ are
the conformations which belong to the previously designed linear
rearrangement path in conformation space.  These conformations are
supposed to form a deep valley in the energy landscape.  For the
set of two target conformations, sequence design was performed in
the paper \cite{2states}.  In that work, the goal was to model
proteins which can fold into two (or more) distinct "native"
conformations, like prions \cite{Prusiner}.  Accordingly, two
target conformations were chosen to be totally dissimilar
(non-overlapping, or weakly overlapping).  In more details, such
design was examined in \cite{Multiple_Funnels}.  In our case, the
problem is almost the opposite. The conformations in question are
very closely related, they can be mutually transformed into one
another in just a few moves. Accordingly, the overlap between
neighboring conformations along the path is very high.  Of course,
as the system walks along its pathway from one end to another, the
overlap decreases, but still remains significant. For example, the
overlap between the conformations at the opposite ends of the path
shown in the Figure \ref{fig:path} is as high as about $75\%$.

The sequence optimization is governed by the following
Hamiltonian:
\begin{equation}
\label{Hdes} {\cal H}_{\rm des}({\rm seq}, \{{\cal
C}_k\})=\sum_{k=1}^{N_{\cal {C}}}{\cal H}({\rm seq}, {\cal C}_k) \
,
\end{equation}
where $N_{\cal C}$ is the total number of conformations  along the
rearrangement pathway.

The question which arises now is this.   How efficient is the
sequence optimization in the case of multiple closely related
target states?  Let us consider the simple situation when only
energies of two target conformations are optimized.  These
conformations, namely ${\cal C}_1$ and ${\cal C}_2$, could be, for
example, the ends of the rearrangement pathway.  How deeply can
their energies be lowered during the sequence design in comparison
with an arbitrary other compact conformation ${\cal C}_{\star}$?

To make this estimate we can calculate the energy of the
conformation ${\cal C}_{\star}$ averaged over the sequence space
\begin{equation}
\langle E_{\star}(seq) \rangle = \frac{ \sum_{seq}P_{seq}^{(0)}
e^{- \left. \left({\cal H}_{\rm des}({\rm seq}, \{{\cal C}_k\})
\right) \right / T_{\rm des} } {\cal H}({\rm seq}, {\cal
C}_{\star})}{ \sum_{seq}P_{seq}^{(0)} e^{- \left. \left({\cal
H}_{\rm des}({\rm seq}, \{{\cal C}_k\}) \right) \right / T_{\rm
des} }}  \ ,
\end{equation}
where $P_{seq}^{(0)}=\prod_{I=1}^N p_{s_I}$ is the probability for
the sequences made randomly from independent monomer species with
occurrence probabilities $p_i$.  The details of similar
calculations are described in the review article \cite{Grosberg1}.
The result for the present case of two target conformations reads
\begin{equation}
\label{Estar} \langle E_{\star}(seq)\rangle=\bar{B}Q^{\star} -
\frac{\delta\hat{B}^2}{T_{\rm des}}[Q^{1, \star}+Q^{2, \star}] \ ,
\end{equation}
where
\begin{eqnarray}
\bar{B}=\sum_{ij} p_i B_{ij} p_j ~~{\rm and}\\
\delta\hat{B}^2=\sum_{ij} p_i (B_{ij}-\bar{B})^2 p_j
\end{eqnarray}
are the mean value and the variance of the interaction matrix,
\begin{equation}
Q^{\star}=\sum_{I<J}^N \Delta({\vec r}_I^{\star}-{\vec
r}_J^{\star})
\end{equation}
is the total number of contacts in the conformation ${\cal
C}_{\star}$ and its overlap with arbitrary  target conformation
${\cal C}_k$ is defined as
\begin{equation}
Q^{k, \star}=\sum_{I<J}^N \Delta({\vec r}_I^{k}-{\vec
r}_J^{k})\Delta({\vec r}_I^{\star}-{\vec r}_J^{\star}) \ .
\end{equation}
As we can see from the expression (\ref{Estar}),  the energy of
the designed sequence in the conformation ${\cal C}_{\star}$
depends on the similarity of this conformation to the target
states.  This similarity is measured by the overlap parameter
$[Q^{1, \star}+Q^{2, \star}]$.  It takes the maximal value for the
a given pair  ${\cal C}_1$, ${\cal C}_2$, when ${\cal C}_{\star}$
coincides with either ${\cal C}_1$ or ${\cal C}_2$.  Usually, if
${\cal C}_{\star}$ lies on the rearrangement pathway, the overlap
parameter $[Q^{1, \star}+Q^{2, \star}]$ takes values close to the
maximum. Therefore, not only energies of the target conformations
are optimized, but conformations between them (e.g.  along the
designed linear path!) are optimized, too.

In principle, sequence design method may also lower the energies
of other states, which are related to the target states, but do
not belong to the rearrangement pathway, and thus are non
desirable for us here.  This is a well known problem, generally
addressed through the "negative design" (see, for instance,
\cite{negative}).  Luckily, in our specific case, conformation
design, as discussed above, helps to address this problem. Indeed,
since there are no allowed conformations at all on the sides of
the pathway, the only possible low energy decoys are structurally
unrelated ones.

Sequence design method employed here may seem to contradict the
results of the recent work \cite{manystates}.  In this work,
authors estimated the maximal possible number of "native" states
which may be "memorized" by the sequence.  They showed that this
number is very limited, it is independent on protein length, and
is fully determined by the alphabet - the number, $q$, of distinct
monomer species (for instance, for the system with $q=20$ monomer
species, there can be not more than four or five "native" states).
In fact, there is no contradiction. The estimate of the work
\cite{manystates} determines the maximal number of unrelated
conformations which can be designed into the sequence.  In our
case, all target states are very closely related, and they in fact
belong to the same potential well, or funnel, in the free energy
landscape.  They, of course, cannot be considered as independent.
As we show below, our sequence design process works successfully
even when the number of states in the rearrangement path is as
large as about ten.

\subsection{Sequence design with energy gradient along the rearrangement path}

So far, we have been discussing the sequence design procedure for
which all the target states ${\cal C}_k$ were equal.  Now,
following Orwell \cite{Orwell}, we shall consider some of the
conformations more equal than others.  Specifically, we should
remember that conformations ${\cal C}_k$ form a one-dimensional
path.  We did it on purpose, and we should use it now.  To be
specific, let us assume that conformations ${\cal C}_k$ are
labeled with index $k$ in a natural way, such that $k$ changes
orderly from $1$ at the one end of the rearrangement pathway to
the maximal value $N_{\cal C}$ at the other end. Then, we shall
require that, say, ${\cal C}_1$ has the lowest energy, ${\cal
C}_2$ has energy a little higher - preferably, by a certain amount
$W$ higher; ${\cal C}_3$ we want to be about $W$ higher in energy
than ${\cal C}_2$, ... , and this continues all the way up to
${\cal C}_{N_{\cal C}}$, which we want to be about $\left( N_{\cal
C} - 1 \right) W$ above ${\cal C}_1$, but still much lower than
all the other conformations.

Why do we want such energy landscape?  First of all, being all
connected in one valley, the conformations ${\cal C}_k$ form
\emph{together} a basin of attraction for folding, or folding
funnel.  Second, when correctly folded and at the bottom of the
funnel, the system can still travel back and forth along the
one-dimensional rearrangement path.  This travel may be either due
to fluctuations or due to some externally applied force.

To achieve this aim we modified the design Hamiltonian
(\ref{Hdes}) in the following way:
\begin{eqnarray}
\label{Hdes2}  && {\cal H}_{\rm des}({\rm seq}, \{{\cal C}_k\})  =
\sum_{k=1}^{N_{\cal {C}}}{\cal H}({\rm seq}, {\cal C}_k)+
\nonumber
\\ & + & \lambda \sum_{k=2}^{N_{\cal {C}}} \left[ {\cal H}({\rm seq}, {\cal
C}_k) - {\cal H}({\rm seq}, {\cal C}_1) - k W \right]^2 \ ,
\end{eqnarray}
where $\lambda$ is the "experimentally" adjusted parameter, and
$W$ is the desired energy gap between neighboring conformations.
The Monte-Carlo optimization ruled by the Hamiltonian
(\ref{Hdes2}) has a bias towards sequences with lower energy in
conformation ${\cal C}_1$ and subsequently higher energies in
conformations ${\cal C}_2,~{\cal C}_3,...$.

Various regimes are possible here depending on the relation
between the design temperature $T_{\rm des}$, real temperature
$T$, and the value of $W$.  Not entering the discussion of all
these regimes, we mention that in what follows we have chosen the
sequence optimization temperature to be $T_{\rm des} = 0.1 \delta
\! B$.

\section{Results}

\subsection{Design of the conformations and the sequence}

To demonstrate the work of our design method, we generated two
sets of conformations of lattice $63$-mer as described above in
the section \ref{switches}.  The linear rearrangement path of the
conformation shown in Figure \ref{fig:path} includes 7
conformations. In another example, shown below in Figure
\ref{fig:substrate}, there are 6 conformations in the path. The
location of the switching elements in these two cases is chosen
differently. Switching elements in the former case are located in
the bulk of the globule, whereas in the latter case the switching
elements are all located near the surface.

For these two sets of target conformations, the sequence
optimization procedure was applied.  The values of parameters were
as follows:   $W=0.3 \delta \! B$, $T_{\rm des} = 0.1 \delta \!
B$, $\lambda=15$, and the interaction matrix was chosen to be
correspond to the independent interactions model (IIM).

\subsection{Folding}

\begin{figure}[ht]
\centerline{\scalebox{0.85} {\includegraphics{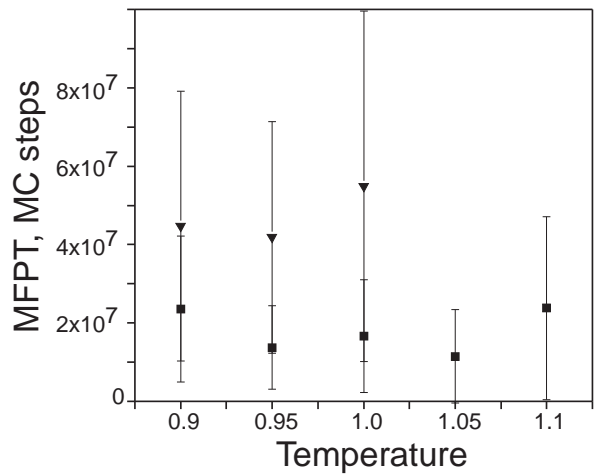}}}
\caption{  The comparison of the folding rates of the two
sequences.  First sequence (squares) is optimized to have unique
ground state , which is the conformation 1 in the Figure 4. Second
sequence (triangles)  is designed according to the approach
described in this paper and has $7$ low energy conformations (all
are shown in the Figure 4. The same IIM matrix with $\langle B
\rangle=-0.025,~~\delta \! B=1.0$ used for the design and folding.
Both sequences were designed at the temperature $T_{\rm
des}=0.1\delta \! B$.}\label{fig:rates}
\end{figure}

First of all, we have to check that the chains can correctly fold
into the conformation as designed.  We used the set of
conformations shown in the Figure 4\ref{fig:path} as target
states.  We compared folding rates for the sequence designed as
proposed in this article and for the control sequence, designed in
a more traditional way, with conformation ${\cal C}_1$ as the only
purported ground state.  All the folding experiments were started
from different random unfolded conformation. The Monte Carlo
simulations  were performed at  temperatures in the range $T=0.7
\div 0.9 T_m$, where $T_m \approx 1.3 \delta \! B$ was the
midpoint temperature.  Mean first passage time (MFPT) at every
temperature was calculated by averaging over $30$ folding runs.
The results are shown in the Figure \ref{fig:rates}.  The folding
times for the sequence which has multiple ground states are
approximately $3$ times longer than for the sequence with the
unique native conformation.  Further inspection suggests that this
happens because the depth of the global minimum for the sequence
with multiple ground states can not be as well optimized as that
for the sequence with unique ground state.  Nevertheless, the
emphasis of our result here is on the good news, not the bad ones:
it is not important that our sequences are slower, it is important
that they are insignificantly slower, \emph{only} by a factor of
about $3$ slower.  That means, they do fold, and their folding
time is of the same order of magnitude.

It is worth to emphasize that during this folding experiments, the
chain was not confined in any restricted volume.  We used volume
restriction in the preliminary stage of this work, to elucidate
the method of conformation design.  Now, as we are done with the
design, we let the polymer do whatever it wants, and the result is
that it folds and spontaneously arrives into the valley where
it has a linear chain of conformations at its disposal.

\subsection{Diffusion along the designed path}

Now, since we have established that our model does fold, we have
to check if it can move along the designed path.  In reality,
conformational relaxation of a functioning protein machine is
triggered by the attachment or detachment of a substrate or other
ligand; we shall consider this in the next section.  Here, we want
to perform the simpler test to see what happens without any
stimuli.  In this situation, we expect our toy protein to move
randomly back and forth along the designed path.  It should be
mostly in the lowest energy state ${\cal C}_{1}$, but since the
energy difference between states along the path $W = 0.3 \delta \!
B$ is only a fraction of thermal energy, bias towards the ${\cal
C}_{1}$ end should be relatively weak, there should be plenty of
fluctuations along the path.  The point to be checked is that the
system, while performing this random walk along the path, should
not open up too frequently, the globule should stay compact.

\begin{figure}[ht]
\centerline{\scalebox{0.85} {\includegraphics{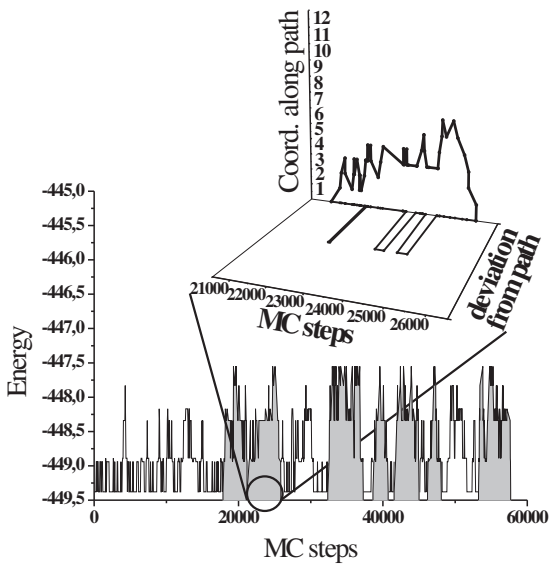}}}
\caption{  A typical MC trajectory of the toy-protein shown in the
Figure \protect\ref{fig:path} at the ground state.  The
dependence of energy on the number of MC steps; (Inset) The dependence
of the conformational coordinates on the number of MC steps.
Coordinate along the linear rearrangement path is shown in the
vertical plane.  The deviations from the linear path conformations
a shown schematically  as  the non-zero values in the horizontal
plane. These deviations are related to the events of unfolding of
some loops on the surface of the molecule.}\label{fig:excursion}
\end{figure}

To examine how this happens in the toy-protein shown in the Figure
\ref{fig:path}, we run a long Monte Carlo simulations at different
temperatures starting from the conformation ${\cal C}_1$.  The
events of passing the conformations $\{{\cal C}_k\}$ of the
pre-designed path are recorded.  Coordinate along the pathway
takes the value $x=k$ if the vacancy position along the path
coincides with conformation ${\cal C}_k$.  A typical "trajectory"
of conformational changes for the first designed toy protein is
shown in Figure \ref{fig:excursion}. The inset of Figure \ref{fig:excursion}
displays the details of one particular passage along the path from
conformation ${\cal C}_1$ to ${\cal C}_{7}$ and back. The events
when conformation changes in a way other than walking along the
path (say, some loop opens on the surface of the globule), are
pictured as the change of the coordinate in the perpendicular
plane.  As one can see, below the midpoint temperature the
toy-protein stays steadily on the rearrangement path and makes
random walks back and forth along it, with very limited
fluctuations in the multitude of transverse directions in the
conformation space.

\begin{figure}[ht]
\centerline{\scalebox{0.85} {\includegraphics{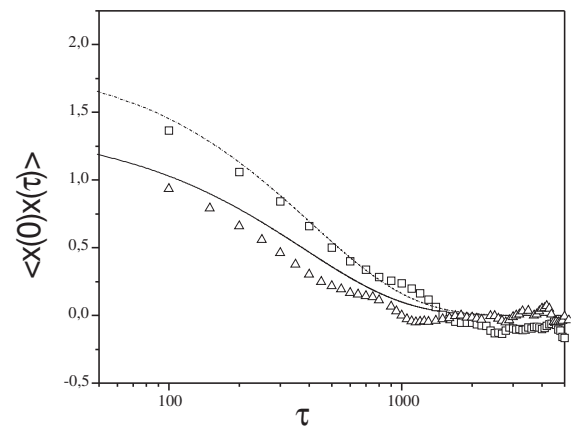}}}
\caption{  The correlation function of the random walk along the
linear path.  The simulation results ($T=0.5\delta B$ - triangles;
$T=0.7\delta B$ - squares) are compared with the curves calculated
for the particle performing one-dimensional walk at the same
potential and at the same temperatures ($T=0.5\delta B$ - solid
line; $T=0.7\delta B$ - dots).}\label{fig:corr_function}
\end{figure}

Thus, what we observe confirms that our toy protein performs
thermal fluctuations in the form of one-dimensional movement along
the designed path.  This is important, because if fluctuations
occur that way, that means, under some different circumstances,
e.g., applied external force, the system can perform also a forced
movement along that same pathway, as it is expected for the
function.  Thus, we need to test carefully that the observed
fluctuations are indeed the random walks effectively in one
dimension.  To address this quantitatively we calculate the
correlation function $\langle x(t)x(t+\tau)\rangle$ for the toy
protein and compare it with that for the artificial auxiliary
random walker which diffuses on the $1D$ discrete set of states
with the same energy profile as for the protein. The good
agreement between the two correlation functions is evident in
Figure \ref{fig:corr_function}.  Hence the designed toy protein
does indeed move along the one-dimensional pathway and in this
sense it remembers its "function."

\subsection{Simulation of the binding/dissociation with the small
molecule}

In the previous section we demonstrated that  toy-protein  can use
the designed rearrangement path, performing a biased, but random
motion. This biased walk can be triggered if the protein molecule
is externally stimulated.  As in the real biochemical world, this
can be most efficiently done by attaching the ligand molecule to
the protein globule.  In this sense, the typical cycle of a
protein work may be described as follows \cite{Blumenfeld}.

We start from the protein molecule in its ground state.  At some
moment, a ligand molecule gets attached to the protein.  With the
ligand attached, protein conformation is no longer the ground
state, some other conformation is now lower in energy, and,
therefore, attachment of ligand initiates the relaxation process
of the globule into the new conformational state, which is the
ground state for the protein-plus-ligand complex.  When this
relaxation is completed (or nearly completed), the new
conformation turns out well suitable for certain chemical (or
other small length scale) changes, the result of which for the
protein is the desorbtion of a ligand initiating again the
relaxation process, this time back into the original ground state.
This type of conformational relaxation processes coupled with the
ligand binding plays is well known for motor proteins
\cite{motor1}, heme-containing proteins \cite{Frauenfelder}, etc.
In this work, following our general strategy, we want to design a
toy ligand with which our toy protein can perform the entire cycle
of its "function."

\begin{figure}[ht]
\centerline{\scalebox{0.85} {\includegraphics{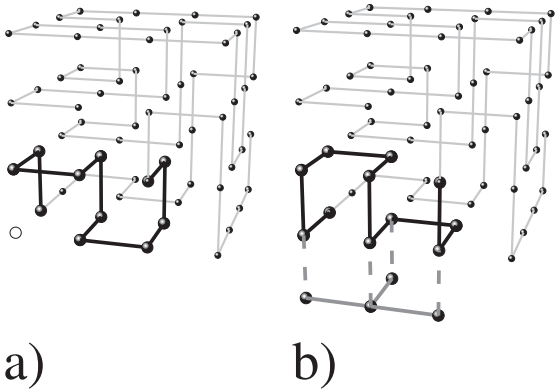}}}
\caption{  Two conformations of the toy-protein representing the
ends of the linear rearrangement path and the small molecule bound
to the active center of the protein (a). The substrate binds when
the protein stays in the conformation ${\cal C}_1$ and unbinds at
the conformation ${\cal C}_6$ (b).}\label{fig:substrate}
\end{figure}

\begin{figure}[ht]
\centerline{\scalebox{0.85} {\includegraphics{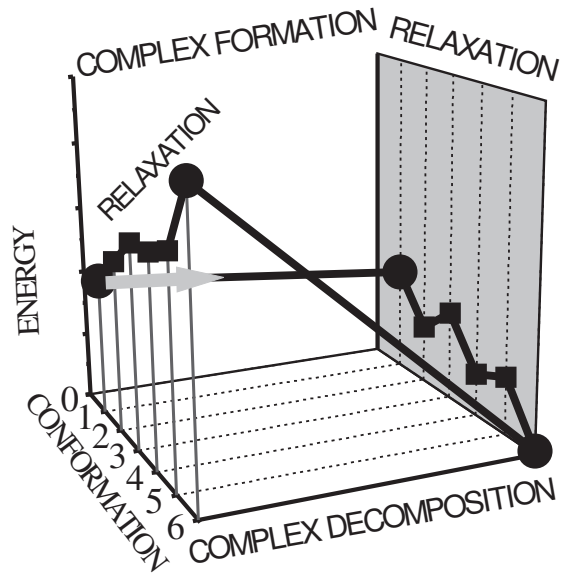}}}
\caption{  The energies of the conformations along the path in the
protein molecule and in the complex.}\label{fig:cycle_energies}
\end{figure}

We designed a toy protein which has binding site and is able to
change its shape.  The initial protein conformation and
conformational changes activated by the adsorption of the toy
ligand are shown in Figure \ref{fig:substrate}.  The toy protein
was designed as explained above. The linear rearrangement path of
the molecule includes $6$ conformations.  The conformation ${\cal
C}_1$ is the lowest energy state for the protein with no ligand.
The energies of interaction of the ligand with the "active center"
of the protein are chosen such that as soon as the complex forms,
the energies of linear path conformations change in comparison
with those in isolated protein molecule.  The profile of the
energy changes along  the path of conformational rearrangements in
the protein-ligand complex is shown in Figure
\ref{fig:cycle_energies}.   When the ligand is attached, the
conformation ${\cal C}_6$ of the path has the lowest energy. Hence
as the small molecule binds to the protein in the state ${\cal
C}_1$, it induces the cascade of conformational changes which
drives the protein to the conformation ${\cal C}_6$. The energy
profile along the path of the conformational changes corresponds
to the case of the suicide inhibitor due to the high barrier for
the dissociation of the ligand in the conformation ${\cal C}_6$.

\begin{figure}[ht]
\centerline{\scalebox{0.85} {\includegraphics{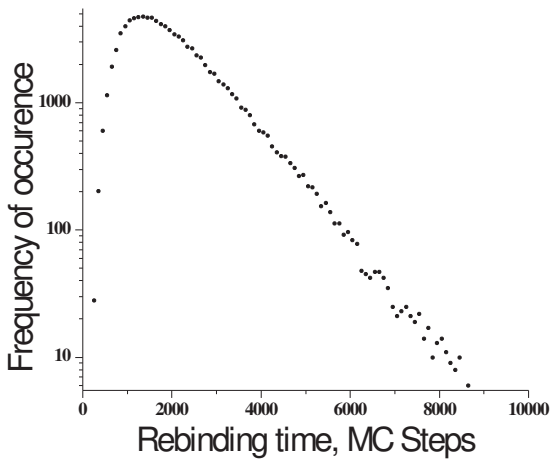}}}
\caption{  The distribution of the relaxation times of the toy
protein in response to the events of ligand binding.  The MC
simulations were performed for the sequence designed in MJ model,
$\langle B \rangle = -1.78,~~ \delta \! B = 1.0,~~T_{\rm des}=0.1
\delta \! B,~~ T=0.4 \delta \! B$.}\label{fig:relax_time}
\end{figure}

Thus, our model exhibits the conformational response to the ligand
binding.  When the ligand binds, it jump-starts the cascade of
orderly occurring conformational transitions ${\cal C}_1
\rightarrow {\cal C}_2 \rightarrow \ldots \rightarrow {\cal C}_6$.
In the computer experiment, we have measured the relaxation times
- the number of Monte Carlo steps in which the toy protein goes
from ${\cal C}_1$ into ${\cal C}_6$ upon ligand binding.  We
repeated this measurement in $100000$ independent MC runs, and
Figure \ref{fig:relax_time} shows the distribution of relaxation
times.  The tail of the distribution demonstrates exponential
decay that is expected for the biased diffusion in one-dimensional
system.

These simulation results are in good agreement with some
experimental data.  For example, Beece et. al. \cite{Frauenfelder}
studied the rebinding of carbon monoxide to myoglobin protein
after dissociation induced by 1 $\mu$s laser impulse.  Such a
short pulse synchronized conformational transformations of many
protein molecules in solution.  Under proper circumstances, the
population of excited protein molecules exhibited exponential
decay, quite similar to our data presented on Figure
\ref{fig:relax_time}.  To reproduce also the multi-exponential or
power law decay on the medium time scale observed in some
experiments \cite{Feng_Gai} will require some modifications of the
present model.

\section{Concluding remarks}

In this study, we designed a toy model which exhibits quite a few
protein-machine-like properties.  First and foremost, it has the
possibility to move like a mechanical system along the
one-dimensional path, which we called rearrangement path. Like
real protein, our model can fold into the valley of conformations
the bottom of which is the rearrangement path.  Thus, it folds,
but not only it just folds, it does so to form a functional,
movable state.  While folded into its low energy valley, our toy
protein can move along the path which is the valley bottom. Making
only few transverse excursions, it performs a biased random walk
along this path.  Attachment and detachment of the ligand can
switch the direction of the bias, thus making toy protein to go
orderly through the closed cycle mimicking protein function.

The usefulness of a toy is the possibility to play.  Playing with
our toy allows to gain some insights into the general concept of
machine-like function of proteins.  One question which we can
discuss is that of storing energy in the deformed globule. Our
model is designed in such a way that the energy of the globule at
one end of the rearrangement path is higher than at the other -
see, e.g, figure \ref{fig:cycle_energies}, left panel.  Can we say
this globule acts as a molecular spring?  To begin with, this
seems to contradict the formulation of the design Hamiltonian
which purports to make energy \emph{linear} in the coordinate $k$
along the path ($k W$), while regular Hookean spring, of course,
should have energy quadratic in deformation.  In fact, this is
unimportant, because design Hamiltonian can be easily modified to
purport quadratic (or any other) energy profile; only discrete
character of the lattice renders such fine tuning of the model
useless.  More importantly, all energies involved in our model are
in fact free energies.  It should be understood that proteins,
although machines, are not heat engines: they function in
essentially isothermic environment.  Therefore, when we say, for
instance, that contact of monomers $i$ and $j$ has energy $\delta
\! B_{ij}$, we should have in mind that pre-averaging has been
performed over a multitude of small scale, rapidly relaxing
degrees of freedom (e.g., $\chi$-angles of the residues side
groups), and then $\delta \! B_{ij}$ is the free energy of the
corresponding contact.  From that point of view, we can say that
at the upper-energy end of the rearrangement path our model
protein stores free energy rather than energy.  In other words, it
acts not so much like a Hookean spring, but rather like a piece of
rubber.  This analogy gets even better if we remember that
linearity of the strain-stress relation and reversibility of
deformation are totally unrelated in the case of rubber.  As in
other mechanical devices, the question of reversibility in our toy
machine is related to its speed: the slower is the rate of
operation, the closer it can approach to the limit of being
reversible.  Of course, when our toy protein undergoes
fluctuations along its designed path, it exchanges energy with the
surrounding heat bath, because the process occurs at the constant
temperature.  When conformational relaxation takes place, because
random walk along the path is strongly biased - then energy
exchange with thermal bath is dominated by the energy transfer
into the bath. However, when we imagine that ligand energy has
driven the enzyme to the high energy end of its path, then this
energy remains available and drives the subsequent conformational
relaxation.  In this delicate sense, our toy can be said to work
like a molecular spring - or, once again, as a piece of rubber. It
goes without saying that this is an overdamped spring (or rubber),
no oscillations are in question in the course of conformational
relaxation.

Our approach is, no doubt, very schematic.  It uses to the full
extent the discrete geometry of moves on the cubic lattice.  This
is the heavy price we have to pay for the tractability of the
model.  In our opinion, the wonderful properties which we found
justify the study of this model.   We hope also that the
fundamental ideas behind our approach will be useful for the more
realistic models, including off-lattice ones.  These fundamental
ideas are the search for a one-dimensional rearrangement pathways
in the compact globule and the design of sequences capable of
folding into a "collective funnel" formed together by all states
belonging to this pathway.  Design of the static protein backbone
conformation \cite{Tang}, as well as sequence design for the
static \cite{seq_des_static} limited flexibility
\cite{seq_des_flex} backbones are all familiar computational
approaches.  It is a challenge to incorporate the search for the
one-dimensional rearrangement pathways into this already difficult
area.

\section{Acknowledgements}

We acknowledge Minnesota Supercomputing Institute whose
computational resources were employed in this study.

AG acknowledges useful discussions and correspondence with
L.~A.~Blumenfeld and A.~S.~Mikhailov.


\begin{thebibliography}{99}
\bibitem{fold_book}E. I. Shakhnovich, R. A. Broglia, G. Tiana (Editors),
Protein Folding, Evolution and Design. In: NATO Sci. Ser., Ser. I,
(2001)
\bibitem{motor} R.D. Vale, T. Funatsu, D.W. Pierce, L. Romberg, Y.
Harada, T. Yanagida, \emph{Nature}  \textbf{380} 451-3 (1996)
%
\bibitem{McClare} C. McClare \emph{J. Theor. Biol.} \textbf{30},
1 (1991).
%
\bibitem{Blum_Russian} L.A. Blumenfeld \emph{Problemy biologicheskoi fiziki (Problems of Biological
Physics)} (\emph{in Russian}). (Moscow: Nauka, 1977) [Translated
into English (Berlin: Springer-Verlag, 1981)]
%
\bibitem{Chernavskii} D. S. Chernavskii, N. M. Chernavskaya {\emph
Protein as a Machine. Biological Macromolecular Constructions}
(Moscow University Publishing, 1999, \emph{in Russian}).
%
\bibitem{Blumenfeld} L. A. Blumenfeld, A. N. Tikhonov \emph{Biophysical
Thermodynamics of Intracellular Processes: Molecular Machines of
the Living Cell} (Springer, NY, 1994).
%
\bibitem{Gruler} H. Gruler et al, \emph{Phys. Rev. E}, \textbf{56},
7116 (1997).
%
\bibitem{Mikhailov} B. Hess et al, \emph{J. Phys. Chem. B},
\textbf{102}, 6273 (1998).
%
\bibitem{database} M. Gerstein, W. Krebs, \emph{Nucleic Acid Res.}
\textbf{26}(18) 4280-4290 (1998)
%
\bibitem{DENSE} E. I. Shakhnovich, \emph{Curr. Op. Struct. Biol.} \textbf{7}(1),  29-40 (1997);
N. D. Socci, J. N. Onuchic, P. G. Wolynes, \emph{Proteins:
Struct., Func. Gen.} \textbf{32}(2), 136-158 (1998); K. A. Dill et
al. \emph{Protein Science} \textbf{4} 561-602 (1995); R. Melin, H.
Li, N. S. Wingreen, C. Tang, \emph{J. Chem. Phys.} \textbf{110}
1252-1262 (1999); A. Maritan, C. Micheletti, A. Trovato, J. R.
Banavar, \emph{Nature}  \textbf{406}(6793) 287-290 (2000)
%
\bibitem{VOIDS} J. Liang, K.A. Dill, \emph{Biophys. J.}, \textbf{
81}(2) 751-766 (2001)
%
\bibitem{fluctuations} H. P. Lu, L. Xun. S. Xie, \emph{Science} \textbf{282} 1877-1882 (1998);
L. Edman, Z. F\"{o}ldes-Papp, S. Wennmalm, R. Rigler \emph{Chem.
Phys.} \textbf{247} 11-22 (1999)
%
\bibitem{trans_coord} R. Du, V. S. Pande, A. Yu. Grosberg, T. Tanaka, E.I.
Shakhnovich, \emph{J. Chem. Phys.}  \textbf{108}(1)  334-350
(1998)
%
\bibitem{DIRECT_CURRENT} A. Yu. Grosberg, Contribution in the book
edited by P. Nielaba and M. Mareshal.
%
\bibitem{Baldwin} R. L. Baldwin %\emph{Matching speed and stability.}
Nature, \textbf{369}, 183-184, 1995;  R. L. Baldwin
% \emph{The nature of protein folding pathways: the classical versus the
%new view.}
J. Biomol. NMR, \textbf{5}, 103-109, 1995.
%
\bibitem{Feng_Gai} F. Gai, K.C. Hasson, J.C. McDonald, P.A. Anfinrud, \emph{
Science}, \textbf{279} 1886-1891 (1998)
%
\bibitem{NATURE} G.B. West, J.H. Brown, B.J. Enquist
%A general model for ontogenetic growth.
\emph{Nature},  \textbf{413}(6856), 628-31, 2001; G.B. West, J.H.
Brown, B.J. Enquist
%The fourth dimension of life: fractal geometry and allometric scaling of organisms.
\emph{Science}, \textbf{284} (5420), 1677-9, 1999; G.B. West, J.H.
Brown, B.J. Enquist
% A general model for the origin of allometric scaling laws in biology.
\emph{Science}, \textbf{276} (5309), 122-6, 1997.
%
\bibitem{VS} P. H. Verdier, W. H. Stockmmayer, \emph{J. Chem. Phys.} \textbf{36} 227 (1962)
%
\bibitem{ERGODICITY} This means, every state can be reached via some finite set of allowed
moves starting from every other state.
%
\bibitem{graph1} J.D. Bryngelson, J.N. Onuchic, N.D. Socci, P.G.
Wolynes, \emph{Proteins} \textbf{21}(3) 167-195 (1995)
%
\bibitem{graph2} M. R. Betancourt, J.N. Onuchic, \emph{J. Chem. Phys.} \textbf{103}(2) 773-787 (1995)
%
\bibitem{graph3} R. Du, V. S. Pande, A. Yu. Grosberg, T. Tanaka, E.I.
Shakhnovich, \emph{J. Chem. Phys.} \textbf{111}(22) 10375-10380
(1999)
%
\bibitem{Foam} R. Du, A. Yu. Grosberg, T. Tanaka, M. Rubinstein, \emph{
Phys. Rev. Lett.}    \textbf{84}(11)  2417-2420 (2000)
%
\bibitem{smallworld} A. Scala, L. A. Nunes Amaral, M. Barth\`{e}l\`{e}my,
\emph{Europhys. Lett.} \textbf{55} (4) 594-600 (2001)
%
\bibitem{percolation} C. Moore, M.E.J. Newman, \emph{Phys. Rev. E}, \textbf{62}(5) 7059-7064 (2000)
%
\bibitem{Conf-des} R. Ramakrishnan, J. F. Pekny, J. M. Caruthers, \emph{J.
Chem. Phys.} \textbf{103}(17) 7592 (1995)
%
\bibitem{seq-des} E.I.
Shakhnovich, A. M. Gutin, \emph{Proc. Natl. Acad. Sci. USA}
\textbf{90} 7195 (1993)
%
\bibitem{Grosberg1} A. Yu. Grosberg, \emph{Physics-Uspekhi} \textbf{40}(2)
125-158 (1997)
%
\bibitem{Yue} K. Yue, K. M. Fiebig, P. D. Thomas, H. S. Chan, E. I. Shakhnovich, and K. A. Dill
% A Test of Lattice Protein Folding Algorithms
\emph{Proc. Nat. Ac. Sci. USA}, \textbf{92}, 325-329 (1995)
%
\bibitem{2states} V. I. Abkevich, A. M. Gutin, E.I.
Shakhnovich,  \emph{Proteins} \textbf{31} 335-344 (1998)
%
\bibitem{IIM} E.I. Shakhnovich, A. M. Gutin, \emph{Biophys. Chem.} \textbf{34}
187 (1989)
%
\bibitem{MJ} S. Miyazawa,  R. Jernigan, \emph{Macromolecules} \textbf{18} 534
(1985)
%
\bibitem{Prusiner} S.B. Prusiner, \emph{Proc. Natl. Acad. Sci.
USA} \textbf{95}, 13363-13383, 1998.
%
\bibitem{Multiple_Funnels}  C.R. Locker, R. Hernandez, \emph{Proc.
Natl. Ac. Sci. USA}, \textbf{98}, 9074-9079, 1997
% A minimalist model protein with multiple folding funnels
%
\bibitem{negative} D. B. Gordon, S. A. Marshall, and S. L. Mayo, \emph{Current
Opinion in Struct. Biol.} \textbf{9} 509-513 (1999);
%Energy Functions for Protein Design.
A. G. Street, and S. L. Mayo \emph{Structure with Folding and
Design}, \textbf{7}, R105-R109 (1999).
%Computational Protein Design.
%
\bibitem{manystates} T. M. A. Fink, R. C. Ball,  \emph{Phys. Rev. Lett.}
\textbf{87}(19)  198103 (2001)
%
\bibitem{motor1} E.P. Sablin, R.J.
Fletterick, \emph{Curr. Op. Struct. Biol.} \textbf{11}(6) 716-724
(2001)
%
\bibitem{Frauenfelder} D. Beece, L. Eisenstein, H. Frauenfelder,
D. Good, M.C. Marden, L. Reinisch, A.H. Reinolds, L.B. Sorensen,
T.K. Yue, \emph{Biochemistry} \textbf{19}(23) 5147-5157 (1980)
%
\bibitem{Orwell} G. Orwell, 1984 \emph{New American Library Classics,
NY, 1990}
%
\bibitem{Tang} J. Miller, C. Zeng, N.S. Wingreen, C. Tang, \emph{Proteins}
\textbf{47} 506512(2002)
%
\bibitem{seq_des_static} B. I. Dahiyat, S. L. Mayo, \emph{Science},
\textbf{278} 82-87 (1997)
%
\bibitem{seq_des_flex} P.B. Harbury, J.J. Plecs, B. Tidor, T. Alber, P.S.
Kim, \emph{Science} \textbf{282} 14621467 (1998)
%
\end{thebibliography}
\end {document}